\documentclass[aip,amsmath,amssymb,reprint]{revtex4-2}

\usepackage{graphicx}
\usepackage{dcolumn}
\usepackage{bm}
\usepackage{placeins}
\usepackage{afterpage}
\usepackage[utf8]{inputenc}
\usepackage[T1]{fontenc}
\usepackage{mathptmx}
\usepackage{etoolbox}
\usepackage{xcolor}

\makeatletter
\newcommand*\emails@list{}%
\def\@email#1#2{%
  \endgroup
  \g@addto@macro\emails@list{\produce@RRAP{*#1\href{mailto:#2}{#2}}}%
}%
\patchcmd{\titleblock@produce}
  {\frontmatter@RRAPformat}
  {\frontmatter@RRAPformat\emails@list\frontmatter@RRAPformat}
  {}{}
\makeatother
\begin{document}


\title[Optimizing Pump Conditions of Parametric Amplifiers for Fast Multiplexed Readout]{Optimizing Pump Conditions of Parametric Amplifiers for Fast Multiplexed Readout of Superconducting Qubits}
\author{Jeongwon Kim}%
\affiliation{Quantum Machines Inc., Tel Aviv, Israel}
\affiliation{IQCC, Tel Aviv, Israel}%
\affiliation{SKKU Advanced Institute of Nanotechnology, Sungkyunkwan University, Suwon 16419, Korea}%

\author{Wei Dai}%
\email[Electronic mail: ]{wei.dai@quantum-machines.co}
\affiliation{Quantum Machines Inc., Tel Aviv, Israel}%

\author{Omrie Ovdat}%
\affiliation{Quantum Machines Inc., Tel Aviv, Israel}
\affiliation{IQCC, Tel Aviv, Israel}%

\author{Akiva Feintuch}%
\affiliation{Quantum Machines Inc., Tel Aviv, Israel}
\affiliation{IQCC, Tel Aviv, Israel}%

\author{Nir Alfasi}%

\affiliation{Quantum Machines Inc., Tel Aviv, Israel}
\affiliation{IQCC, Tel Aviv, Israel}%

\author{Yonuk Chong}
\email[Electronic mail: ]{yonuk@skku.edu}
\affiliation{SKKU Advanced Institute of Nanotechnology, Sungkyunkwan University, Suwon 16419, Korea}
\affiliation{Department of Quantum Information Engineering, Sungkyunkwan University, Suwon 16419, Korea}

\date{\today}

\begin{abstract}
Low-noise parametric amplifiers are widely used as the first-stage amplifier in qubit readout chains. 
The performance of parametric amplifiers depends sensitively on the choice of the pump condition. 
We propose a strategy for determining the pump condition that is tailored for fast multiplexed readout. 
Choosing the amplifier pump to maximize the signal-to-noise ratio (SNR) improvement at the readout frequency of the limiting qubit---the qubit that requires the longest readout time to reach a target SNR---minimizes the total multiplexed readout time. 
We demonstrate our pump calibration strategy experimentally on a five-qubit multiplexed readout chain with a traveling-wave parametric amplifier. 
Using our strategy, we reduce the multiplexed readout time by 320\,ns compared to optimizing the average SNR improvement on all qubits, without degrading the target SNR for any qubit.

\end{abstract}

\maketitle

\section{Introduction}

Qubit readout is a ubiquitous operation in quantum computing, from end-of-the-line state measurements to measurement-based 
initialization~\cite{johnson2012heralded, riste2012initialization} and teleportation~\cite{Steffen2013teleportation, Qiu2025teleportation} protocols. 
Most notably, quantum error correction (QEC) ~\cite{ofek_qec,krinner2022realizing,sivak_gkp,google_qec,ni_qec} relies critically on repeated mid-circuit readouts. 
In order to achieve high readout fidelity, the readout operation has to be fast relative to the qubit coherence time. 
The readout speed also limits the achievable cycle rate in repeated sequences. 
Moreover, while some qubits are being measured, other qubits idle and accumulate decoherence for the entire circuit. 
Therefore, minimizing the readout time is a system-level objective along with optimizing the readout fidelity. 

A key enabling technology for fast, high fidelity readout of superconducting qubits is low-noise parametric amplifiers, including resonant Josephson parametric amplifiers (JPAs)~\cite{Bergeal2010_JPC_exp,Hatridge2011_JPA,lin2013single,Frattini2018SPA} and traveling-wave parametric amplifiers (TWPAs)~\cite{o2014resonant,macklin2015,white2015,planat2020photonic,gaydamachenko2025rf}. 
These devices can operate near the quantum limit of added noise~\cite{clerk2010introduction}. Therefore, they are commonly employed as the first-stage amplifier in a cascaded amplifier chain to boost the measurement SNR within short readout time. 
In particular, TWPAs or impedance-matched JPAs (IMPAs)~\cite{Roy2015broadband,White2023readout,Joshi2025lumped,Sun2025merged} featuring broadband gain profile and high saturation power are suitable for frequency-multiplexed readout~\cite{jerger2012frequency, jeffrey2014fast,heinsoo2018rapid,
spring2025fast,castellanos2025measurable}, which is essential for scalable superconducting quantum processing units (QPUs). 
Multiplexed readout of multiple qubits on one output line reduces cryostat RF wiring and bulky microwave components, thereby significantly easing space constraints and heat loads on the mixing chamber~\cite{krinner2019engineering}. 

Achieving fast, high-fidelity multiplexed readout requires low-noise amplification of signals across all readout frequencies simultaneously. 
However, meeting this demand can be nontrivial, as the gain and noise profiles of a TWPA or IMPA can manifest strong frequency dependence due to environmental impedance mismatch~\cite{Mutus2014strong,White2023readout,Kern2023reflection, gaydamachenko2025rf,Sun2025merged, nilsson2024small}. 
Moreover, such impedance mismatch can be modified by the pump condition~\cite{Nilsson2024Peripheral,Dai2025optimizing, zhao2021quantum}. As a result, the multiplexed readout SNR for different qubits can be sensitive to the choice of pump frequency and power; certain pump conditions can improve SNR for some qubits while degrading others. 
Conventional procedure of tuning up the pump for a parametric amplifier targets a nominal gain specification, or the average SNR improvement within the whole band~\cite{nilsson2024small}. 
Such procedure does not account for ripples in the gain profile, and is segregated from bring-up of the multiplexed readout itself~\cite{Sank2025system}. 

In this work, we present a systematic methodology for optimizing the parametric amplifier pump condition to minimize the multiplexed readout time without degrading SNR. 
As all readout channels are acquired simultaneously, the duration of the multiplexed readout operation is set by the slowest readout channel. 
By tuning the amplifier pump to improve this slowest readout channel instead of the average readout channel, we demonstrate a reduction of multiplexed readout time without sacrificing the SNR of any individual qubit. 
The strategy we introduce integrates the bring-up of amplifier pump and readout pulses, and is applicable to a broad range of multiplexed readout architectures. 

\section{\label{methods}Methods}
\afterpage{%
  \clearpage
  \begin{figure}[!t]
    \centering
    \includegraphics[width=8.5cm]{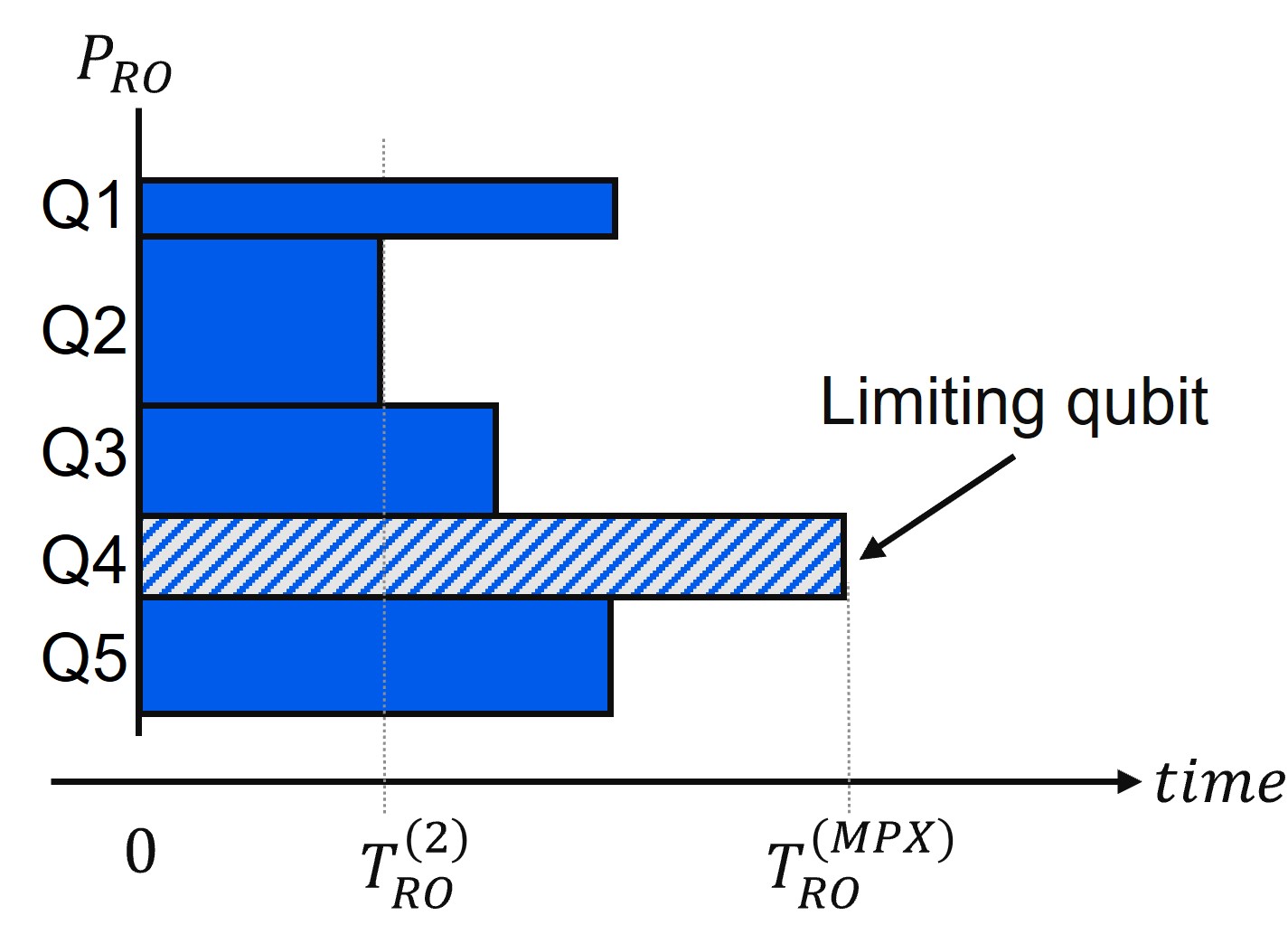}
    \caption{\label{fig:mtplx} Schematic of multiplexed readout pulses for five qubits. All readout tones are applied simultaneously at time 0 on a shared feedline, but the readout amplitude $P_{\mathrm{RO}}$ (pulse height) and the readout time $T_{\mathrm{RO}}^{(i)}$ required for each qubit to reach a target SNR can differ. In this example, Q4 requires the longest readout time, making Q4 the limiting qubit for the multiplexed readout. The total multiplexed readout time $T_{\mathrm{RO}}^{(\mathrm{MPX})}$ is set by that of the limiting qubit.}
  \end{figure}
}%

\subsection{\label{sec:MPXRO}Multiplexed readout time and limiting qubit}

We demonstrate multiplexed readout of five tunable superconducting transmon qubits using distinct, frequency-separated readout resonators coupled to a common feedline, as described in Appendix~A. 
A multi-tone readout pulse addresses all resonators in parallel. 
During readout, each tone acquires a qubit-state-dependent phase shift. 
The combined output signals are amplified by a measurement chain with a TWPA as the first stage amplifier, and demodulated and digitized for state discrimination individually. 

We quantify single-shot readout performance using the signal-to-noise ratio (SNR)~\cite{krantz2019quantum},
\begin{equation}
\mathrm{SNR}=\frac{\mu_e-\mu_g}{\sigma_e+\sigma_g},
\label{eq:snr}
\end{equation}
where $\mu_g$ ($\mu_e$) and $\sigma_g$ ($\sigma_e$) are the mean and standard deviation of the single-shot readout distributions for the ground (excited) state.

The readout SNR for each qubit depends on both fixed device parameters---the resonator linewidth $\kappa$, the dispersive coupling strength $\chi$---and adjustable readout pulse parameters, i.e. readout amplitude and readout time. 
The five qubit--resonator pairs have different parameters. The amplitude of each readout tone $P_{\mathrm{RO}}$ is determined independently such that it does not activate readout-induced leakage~\cite{Hazra2025benchmarking} on the corresponding qubit. 
Consequently, each qubit has a different minimum readout time $T_{\mathrm{RO}}^{(i)}$ required to reach a chosen target SNR (here $\mathrm{SNR}=3$), as schematized in Fig.~\ref{fig:mtplx}. 
As all tones start at the same time, the multiplexed readout operation is completed only after the slowest channel reaches its target SNR. 
Therefore, we define the total multiplexed readout time as $T_{\mathrm{RO}}^{(\mathrm{MPX})}=\max_i T_{\mathrm{RO}}^{(i)}$. 
We refer to the qubit that sets this maximum as the \emph{limiting qubit}, which in our example is Q4. 
Increasing the acquisition time increases SNR, but also increases the probability of energy relaxation ($T_1$) events during the measurement. This trade-off motivates choosing the target SNR that optimizes the readout fidelity, which depends on the qubit $T_1$.  
Using the standard dispersive-readout scaling $\mathrm{SNR}\propto\sqrt{T_{\mathrm{RO}}}$~\cite{clerk2010introduction} in Appendix~B, increasing $T_{\mathrm{RO}}$ initially improves fidelity by increasing SNR, but for finite $T_1$ it eventually yields diminishing returns as relaxation during the readout becomes dominant. Since our five qubits have an average $T_1\approx 30\,\mu\mathrm{s}$, we set the target to $\mathrm{SNR}=3$ and then focus on minimizing $T_{\mathrm{RO}}$ rather than further increasing SNR.

We determine the readout duration for each qubit using the following procedure:
\begin{enumerate}
  \item Sweep the readout pulse amplitude (below a critical strength that starts to induce leakage events) and duration, where we have used square pulses and keep the readout time same as the duration of the pulse. 
  \item For each readout pulse, extract the SNR by fitting the state-dependent readout histograms to a double-Gaussian model.
  \item Among the pulses that achieve $\mathrm{SNR}=3$, select the shortest duration and define it as $T_{\mathrm{RO}}^{(i)}$ for that qubit.
\end{enumerate}
Having identified the limiting-qubit, we then tune the TWPA pump operating point to preferentially improve that channel and thereby reduce $T_{\mathrm{RO}}^{(\mathrm{MPX})}$ (Sec.~\ref{sec:TWPAPump}).

\subsection{\label{sec:TWPAPump} TWPA pump optimization}

\begin{figure*}[!t]
  \centering
  \includegraphics[width=17cm]{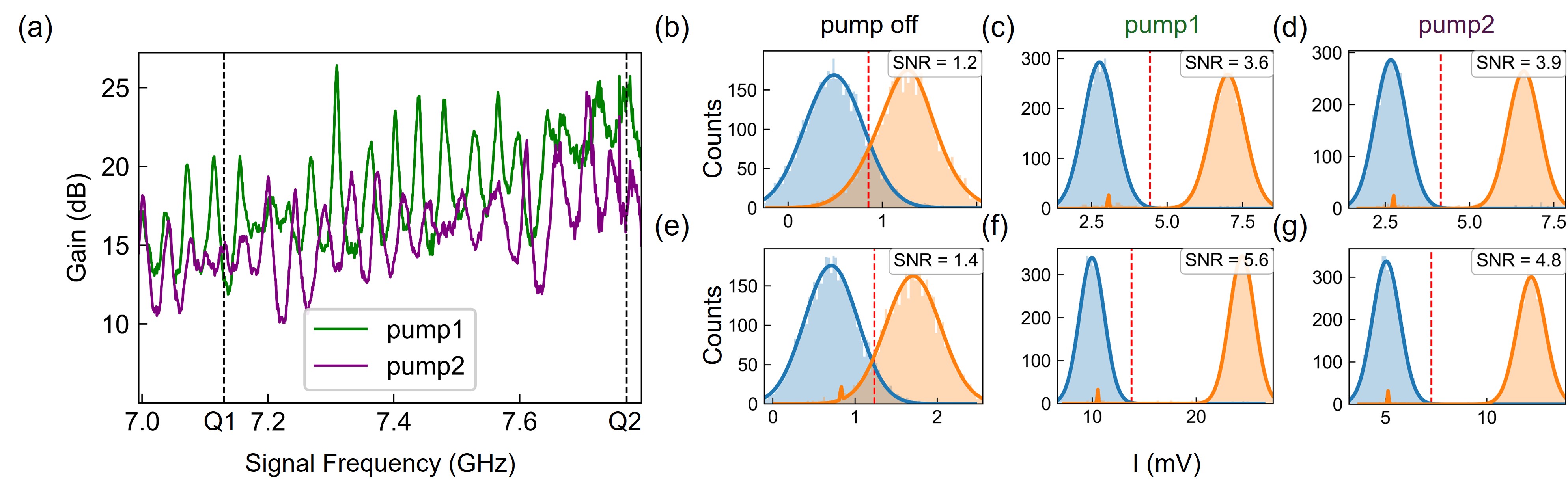}
  \caption{\label{fig:pump}Effect of TWPA pump conditions on readout performance. (a) TWPA gain profiles under two different pump conditions: pump 1 ($f_p=8.428\,\mathrm{GHz}$, $P_p=-85.9\,\mathrm{dBm}$ shown in green) and pump 2 ($f_p=8.382\,\mathrm{GHz}$, $P_p=-85.7\,\mathrm{dBm}$ shown in purple) respectively. 
  The two black dashed lines mark the readout frequencies of two qubits: Q1 (7.12\,GHz) and Q2 (7.77\,GHz). 
  (b--d) Gaussian fits to single-shot readout histograms of Q1 with the TWPA pump off, under pump1, and under pump2, respectively, using the same $1\,\mu\mathrm{s}$ readout pulse. The corresponding SNR (Eq.~1) is annotated in the upper-right corner of each panel; blue (orange) corresponds to the ground (excited) state, and the red dashed line indicates the discrimination threshold in the in-phase quadrature $I$. 
  (e--g) Same as (b--d) for Q2. 
  The data show that the TWPA response depends on both signal frequency and pump condition: at a fixed pump, different qubits experience different gain at their readout frequencies, and for a given qubit the SNR improvement depends on the pump condition (here, Q1 benefits more from pump2, whereas Q2 benefits more from pump1).}
\end{figure*}

We characterize the TWPA performance in terms of gain and SNR improvement ($\Delta\mathrm{SNR}$)~\cite{nilsson2024small} , reporting both metrics in decibels. Here, $\mathrm{SNR}$ (Eq.~\ref{eq:snr}) refers to the single-shot readout SNR for a given qubit, whereas $\Delta\mathrm{SNR}$ quantifies the TWPA-induced improvement in that SNR relative to the pump-off baseline. For each readout resonator, we evaluate both metrics over a 1\,MHz analysis band centered at the readout frequency. We sample $n$ discrete frequency points within this band and average the measured signal and noise voltage amplitudes across those points. 
The gain is
\begin{equation}
\mathrm{Gain}=20\log_{10}\!\left(\frac{\sum_{f_i=1}^{n} S_{\mathrm{on}}^{f_i}}{\sum_{f_i=1}^{n} S_{\mathrm{off}}^{f_i}}\right),
\label{eq:gain}
\end{equation}
where $S_{\mathrm{on}}^{f_i}$ and $S_{\mathrm{off}}^{f_i}$ denote the measured signal voltage amplitude at frequency point $f_i$ with the pump turned on and off, respectively.

We define the SNR improvement $\Delta\mathrm{SNR}$, as the difference between the band-averaged SNR with the pump turned on and the band-averaged SNR with the pump turned off,
\begin{equation}
\Delta\mathrm{SNR}=20\log_{10}\!\left(\frac{\sum_{f_i=1}^{n} S_{\mathrm{on}}^{f_i}}{\sum_{f_i=1}^{n} N_{\mathrm{on}}^{f_i}}\right)-20\log_{10}\!\left(\frac{\sum_{f_i=1}^{n} S_{\mathrm{off}}^{f_i}}{\sum_{f_i=1}^{n} N_{\mathrm{off}}^{f_i}}\right),
\label{eq:dsnr}
\end{equation}
where $N_{\mathrm{on}}^{f_i}$ and $N_{\mathrm{off}}^{f_i}$ are the measured noise voltage amplitudes at frequency $f_i$ with the pump on and off, respectively.

\begin{figure}[t]
  \centering
  \includegraphics[width=8.5cm]{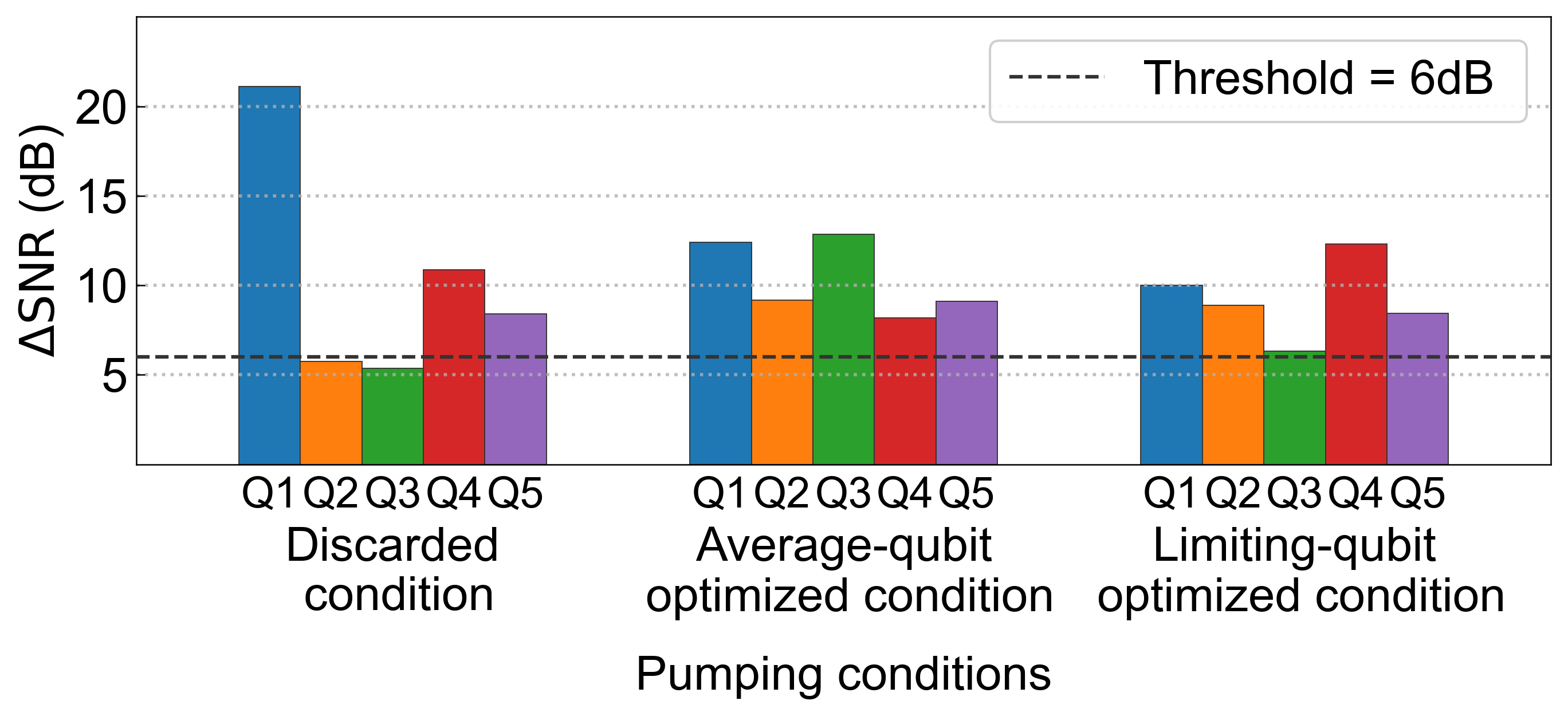}
  \caption{\label{fig:pumpstrategy}$\Delta\mathrm{SNR}$ for each qubit under three representative pump conditions. The black dashed line indicates the minimum acceptable $\Delta\mathrm{SNR}$ threshold (6\,dB in this experiment). The discarded condition (8.244\,GHz, $-87.4$\,dBm) maximizes the average $\Delta\mathrm{SNR}$ across all pump conditions, but it is rejected because Q2 and Q3 fall below the 6\,dB threshold. The discarded condition yields an average $\Delta\mathrm{SNR}$ of 14.93\,dB. The average-qubit-optimized condition (8.311\,GHz, $-86.9$\,dBm) is the feasible pump condition (all qubits above threshold) that maximizes the average $\Delta\mathrm{SNR}$, giving 10.77\,dB. The limiting-qubit-optimized condition (8.328\,GHz, $-87.5$\,dBm) is the feasible setting that maximizes $\Delta\mathrm{SNR}$ for the limiting qubit (Q4), with an average $\Delta\mathrm{SNR}$ of 9.64\,dB.}
\end{figure}

Both gain and $\Delta\mathrm{SNR}$ are dependent on the signal frequency, as shown in Fig.~\ref{fig:pump}(a) and Fig.~\ref{fig:pumpstrategy}. 
Ideally, the TWPA response would be flat across a large bandwidth, but in practice it exhibits gain ripples due to the impedance mismatch and other device nonidealities.
Moreover, the ripple pattern varies under different pump operating points. As a result, changing the pump frequency and power reshapes the gain profile and changes the gain (and $\Delta\mathrm{SNR}$) at a given readout frequency. 
This is apparent at the readout frequencies of Q1 and Q2, shown as black dashed lines, where pump~1 and pump~2 yield different gains in Fig.~\ref{fig:pump}(a). Consistent with this, the single-shot histograms show that the same pump condition can improve the SNR by different amounts for different qubits (Fig.~\ref{fig:pump}(b--g)). 
These observations motivate the pump-selection strategy summarized in Fig.~\ref{fig:pumpstrategy} and described below: we enforce a per-qubit minimum-performance constraint while prioritizing improvement of the limiting-qubit channel under multiplexed readout.

We calibrate the pump as follows. First, we sweep the TWPA pump frequency and power and evaluate the gain and $\Delta\mathrm{SNR}$ at each qubit's readout frequency. We then impose a minimum $\Delta\mathrm{SNR}$ threshold that all qubits must satisfy (6\,dB in this experiment), excluding pump conditions that improve the average performance but degrade one or more channels below threshold, as illustrated by the discarded condition in Fig.~\ref{fig:pumpstrategy}. We compute the average $\Delta\mathrm{SNR}$ across qubits by converting each qubit's $\Delta\mathrm{SNR}$ from dB to linear units, averaging in linear units, and converting back to dB, since dB is a logarithmic scale and therefore cannot be averaged by directly summing dB values and dividing by the number of qubits. Among the feasible pump conditions, we select (i) the operating point that maximizes the average $\Delta\mathrm{SNR}$ across qubits (average-qubit-optimized) and (ii) the operating point that maximizes $\Delta\mathrm{SNR}$ for the limiting qubit (limiting-qubit-optimized). We then compare these operating points in terms of the resulting multiplexed readout time.

\section{\label{results}Results}
\begin{figure*}[t]
\includegraphics[width=17cm]{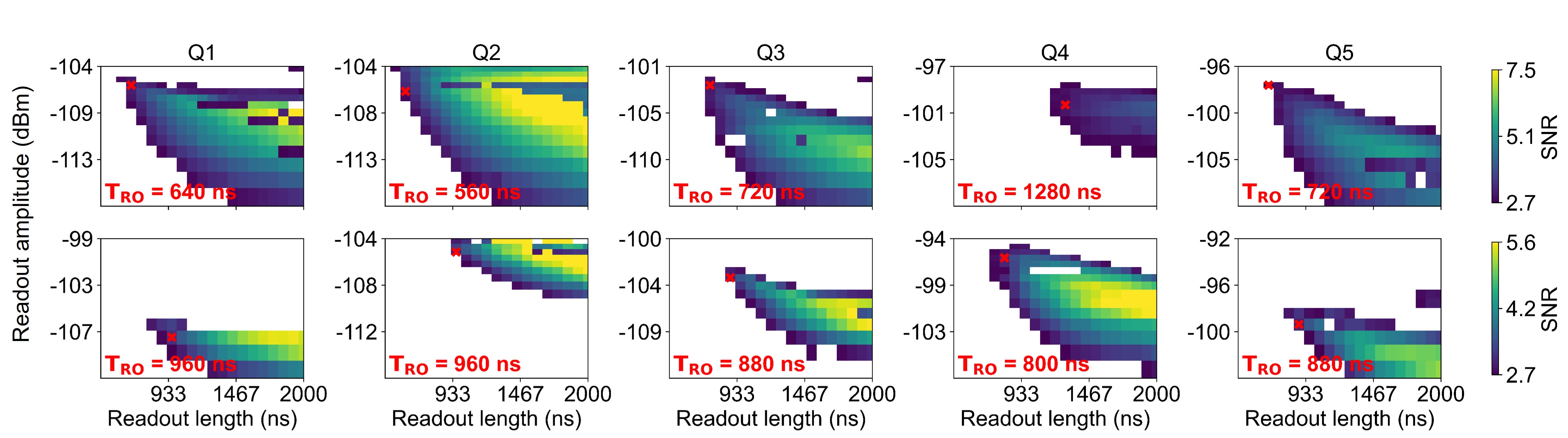}
\caption{\label{fig:result}Readout time required to reach $\mathrm{SNR}=3$ for each qubit under two TWPA pump conditions. Each panel shows the measured SNR as a function of readout amplitude and readout time, and the red cross marker indicates the shortest pulse that achieves $\mathrm{SNR}=3$ as described in Sec.~\ref{sec:MPXRO}. White regions correspond to $\mathrm{SNR}<2.7$. Top row: average-qubit-optimized pump ($f_p=8.311\,\mathrm{GHz}$, $P_p=-86.9\,\mathrm{dBm}$). Q4 requires the longest readout time of 1280\,ns and therefore sets the multiplexed readout time $T_{\mathrm{RO}}^{(\mathrm{MPX})}=1280\,\mathrm{ns}$. Bottom row: limiting-qubit-optimized pump ($f_p=8.328\,\mathrm{GHz}$, $P_p=-87.5\,\mathrm{dBm}$), chosen to maximize $\Delta\mathrm{SNR}$ for Q4 while satisfying the minimum-threshold constraints for all qubits. Q4 shortens by 480\,ns to 800\,ns. The multiplexed readout time decreases from 1280\,ns to 960\,ns, a 320\,ns reduction, and is then set by Q1 and Q2.}
\end{figure*}

Figure~\ref{fig:pumpstrategy} shows how the TWPA pump condition reshapes the per-qubit performance profile and the selected pump points for the following experiment. We first exclude pump conditions that violate a minimum per-qubit requirement, $\Delta\mathrm{SNR}\ge 6\,\mathrm{dB}$. This removes ``ripply'' operating points that maximize the average $\Delta\mathrm{SNR}$ but leave one or more channels below threshold, as illustrated by the discarded condition in Fig.~\ref{fig:pumpstrategy}. Among the remaining feasible conditions, we select the pump that maximizes the average $\Delta\mathrm{SNR}$ across qubits and use it as a reference operating point, which we refer to as average-qubit-optimized.

Using this reference pump, we determine the shortest readout pulse that reaches $\mathrm{SNR}=3$ for each qubit following the procedure in Sec.~\ref{sec:MPXRO}. The red cross marker in Fig.~\ref{fig:result}, top row, denotes the shortest pulse achieving $\mathrm{SNR}=3$ for each qubit. The corresponding $\Delta\mathrm{SNR}$ values of the TWPA, readout times, and fidelity are listed in Table~\ref{tab:twpa}. Under this average-optimized pump, Q4 requires the longest readout time of 1280\,ns and therefore acts as the limiting qubit, setting the multiplexed readout time $T_{\mathrm{RO}}^{(\mathrm{MPX})}=1280\,\mathrm{ns}$.

We then retune the pump to prioritize the limiting qubit Q4. 
Specifically, we choose the feasible pump condition that maximizes $\Delta\mathrm{SNR}$ at Q4's readout frequency, which we refer to as limiting-qubit-optimized, while still enforcing $\Delta\mathrm{SNR}\ge 6\,\mathrm{dB}$ for all qubits. 
As shown in Fig.~\ref{fig:result}, bottom row, and summarized in Table~\ref{tab:twpa}, this increases Q4's $\Delta\mathrm{SNR}$ and reduces its required readout time from 1280\,ns to 800\,ns. 
This improvement comes at the cost of reduced $\Delta\mathrm{SNR}$ for most other qubits, which increases their required readout times to reach the same target $\mathrm{SNR}=3$. 
Under the limiting-qubit-optimized pump, readout of Q1 and Q2 become the new slowest channels, and sets the multiplexed readout time to $T_{\mathrm{RO}}^{(\mathrm{MPX})}=960\,\mathrm{ns}$.
Because we fix the target to $\mathrm{SNR}=3$ for all qubits, the assignment error due to finite SNR is the same between the two pump conditions. The remaining fidelity differences primarily reflect $T_1$ events during readout and other experimental nonidealities such as crosstalk. 
The mean single-shot readout fidelity across the five qubits is 95.45\% for the average-qubit-optimized pump condition and 95.32\% for the limiting-qubit-optimized pump condition, indicating no significant change in average readout fidelity.

The key point is that multiplexed readout time is set by the slowest channel. Therefore, a pump condition that shortens the readout time for the limiting qubit can reduce $T_{\mathrm{RO}}^{(\mathrm{MPX})}$ even if it slightly worsens other channels, provided those channels remain faster than the original slowest channel.

\section{\label{conclusion}Conclusion}
We propose and demonstrate a strategy for selecting the TWPA pump operating point for frequency-multiplexed qubit readout that directly targets the experimentally relevant figure of merit: the total multiplexed readout time $T_{\mathrm{RO}}^{(\mathrm{MPX})}$. 
After identifying the limiting qubit in a multiplexed readout setup, we retune the TWPA pump to preferentially increase $\Delta\mathrm{SNR}$ at the readout frequency of the limiting qubit while enforcing a minimum $\Delta\mathrm{SNR}$ constraint across all qubits. 
This limiting-qubit-optimized pump condition prioritizes the bottleneck channel rather than targeting average performance across all channels. 
As a demonstration, we achieved a 25\% reduction of $T_{\mathrm{RO}}^{(\mathrm{MPX})}$, from 1280\,ns to 960\,ns, on a five-qubit multiplexed line. 
Although the experiment in this work is done with a TWPA, our strategy applies directly to IMPA as well. 

We see the strategy from this work as a practical technique to mitigate experimental nonidealities such as gain ripples, offering a path towards improvement without any hardware change. 
More broadly, this bottleneck-guided tuning strategy provides a route for system-level optimization as multiplexed readout continues to scale. 
Future work will involve further optimization of amplifier pump accounting for gain compression, and shaping of readout pulses to minimize the readout resonator population time.

\begin{table}[!b]
\caption{\label{tab:twpa}Multiplexed readout performance under two TWPA pump operating points (average-qubit-optimized and limiting-qubit-optimized). For each qubit, we list the $\Delta\mathrm{SNR}$, the minimum readout duration $T_{\rm RO}$ required to reach the target $\mathrm{SNR}=3$, and the corresponding single-shot readout fidelity defined. The last column reports the total multiplexed readout time $T_{\rm RO}^{(MPX)}$.}
\centering
\setlength{\tabcolsep}{2.6pt}
\renewcommand{\arraystretch}{1.05}
\begin{ruledtabular}
\begin{tabular}{l|ccccc|c}
 & Q1 & Q2 & Q3 & Q4 & Q5 & $T_{\rm RO}^{\rm (MPX)}$\\
\hline
\multicolumn{7}{c}{\textbf{average-qubit-optimized}}\\
\hline
$\Delta\mathrm{SNR}$ [dB] & 12.4 & 9.17 & 12.86 & 8.18 & 9.10 & \\
$T_{\rm RO}$ [ns]        & 560  & 560  & 720   & 1280 & 640 & 1280\\
Fidelity [\%]            & 90.05 & 96.73 & 96.23 & 97.27 & 96.95 & \\
\hline
\multicolumn{7}{c}{\textbf{limiting-qubit-optimized}}\\
\hline
$\Delta\mathrm{SNR}$ [dB] & 10.0 & 8.89 & 6.34  & 12.30 & 8.43 & \\
$T_{\rm RO}$ [ns]        & 960  & 960  & 880   & 800  & 880  & 960\\
Fidelity [\%]            & 93.86 & 96.05 & 96.36 & 96.14 & 94.18 & \\
\end{tabular}
\end{ruledtabular}
\end{table}

\clearpage
\appendix

\section{MEASUREMENT SETUP}

\FloatBarrier
\begin{figure}[!ht]
  \centering
  \includegraphics[width=\columnwidth]{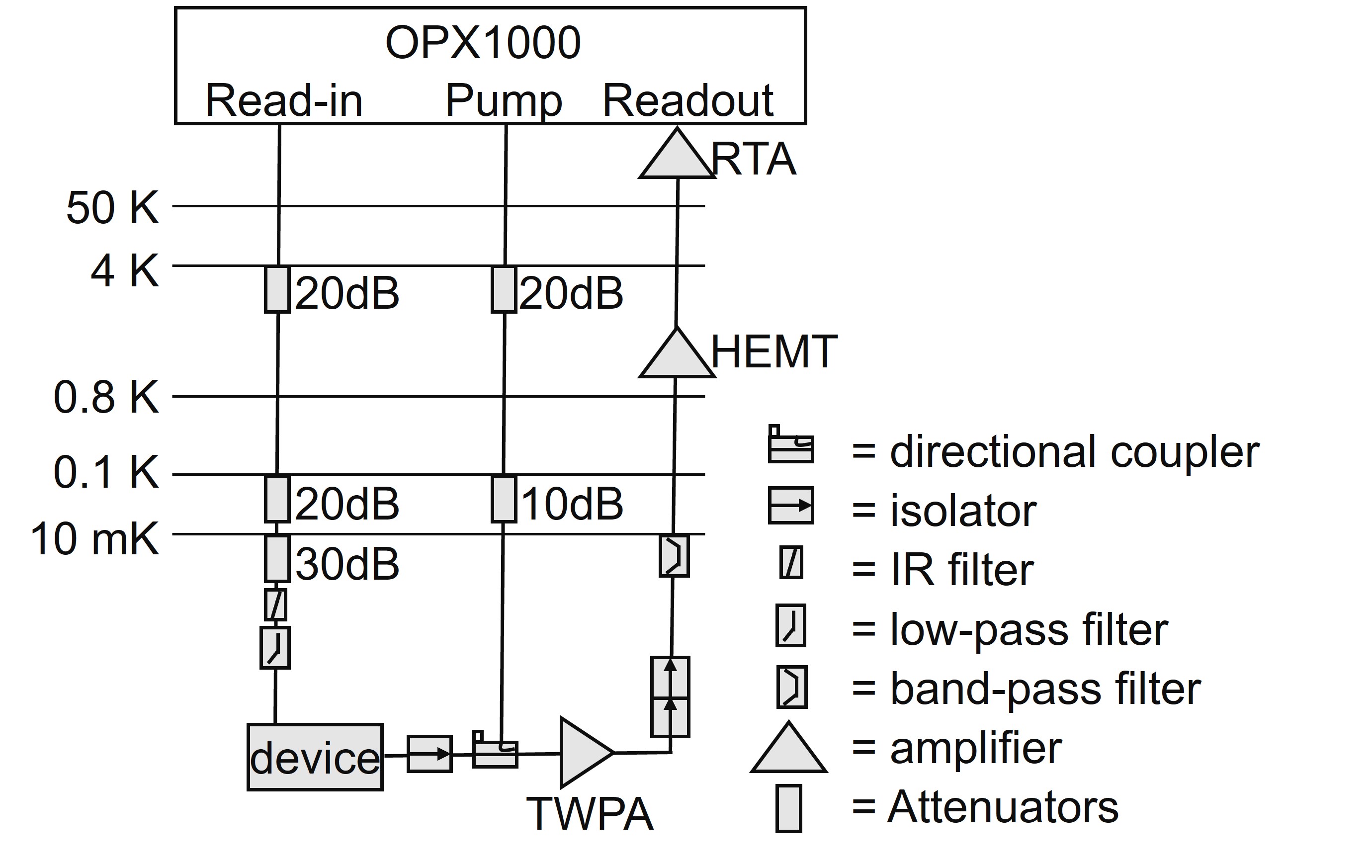}
  \caption{Simplified schematic of the measurement setup used for frequency-multiplexed qubit readout. Readout tones are sent to the device, the transmitted signals are amplified by a TWPA followed by a 4\,K HEMT and room-temperature amplifiers, and the signals are then downconverted and digitized for demodulation and state discrimination.}
  \label{fig:setup}
\end{figure}

All measurements presented in this work were performed at the Israeli Quantum Computing Center (IQCC). Figure~\ref{fig:setup} summarizes the cryogenic measurement chain. We use a Quantum Machines OPX1000 MW-FEM to generate and process the qubit drive tones, multiplexed readout tones, and the TWPA pump, and an OPX1000 LF-FEM to provide the qubit flux-bias signals. The device is a QuantWare Contralto D 21-qubit chip together with a QuantWare Crescendo TWPA. From the 21-qubit array, we select five qubits coupled to a common readout feedline. Each qubit has its own dedicated drive line and flux-bias line, not shown.

The read-in line includes approximately 70\,dB of attenuation and cryogenic filtering, including an IR filter and a 10 GHz low-pass filter, to suppress thermal noise from higher-temperature stages. The TWPA pump is injected through a directional coupler with effective coupling and attenuation of about 20\,dB at the pump port and is kept on during readout. Isolators are placed before and after the TWPA. The isolator between the device and the coupler protects the device from reflected pump power, while the isolator after the TWPA suppresses back-propagating noise from the 4\,K HEMT. The output chain consists of the TWPA, a 4\,K HEMT, and room-temperature amplifiers with about 100\,dB total gain. 

\FloatBarrier

\section{Target SNR}
Readout fidelity in our experiment is defined as the average assignment probability
\begin{equation}
\label{eq:assign_fidelity_def}
\mathrm{Fidelity}=\tfrac{1}{2}\left[P(g|g)+P(e|e)\right],
\end{equation}
where $P(g|g)$ [$P(e|e)$] is the probability to assign the outcome $g$ [$e$] when the qubit is prepared in $g$ [$e$].

To select a target SNR, we use the approximate dispersive-readout model including relaxation derived in Ref.~\cite{kam2024fast},
\begin{equation}
\label{eq:assign_fidelity_model}
\mathrm{Fidelity}=\exp\!\left(-\frac{T_{\mathrm{RO}}}{2T_1}\right)\,
\operatorname{erf}\!\left(\frac{\mathrm{SNR}}{\sqrt{2}}\right),
\end{equation}
which assumes discrimination error arising from two identical Gaussian distributions and $T_1$ error during the readout act independently of each other. Equation~\eqref{eq:assign_fidelity_model} is a justified approximation in the regime of $T_{\mathrm{RO}}\ll T_1$.

\begin{figure}[!ht]
  \centering
  \includegraphics[width=\columnwidth]{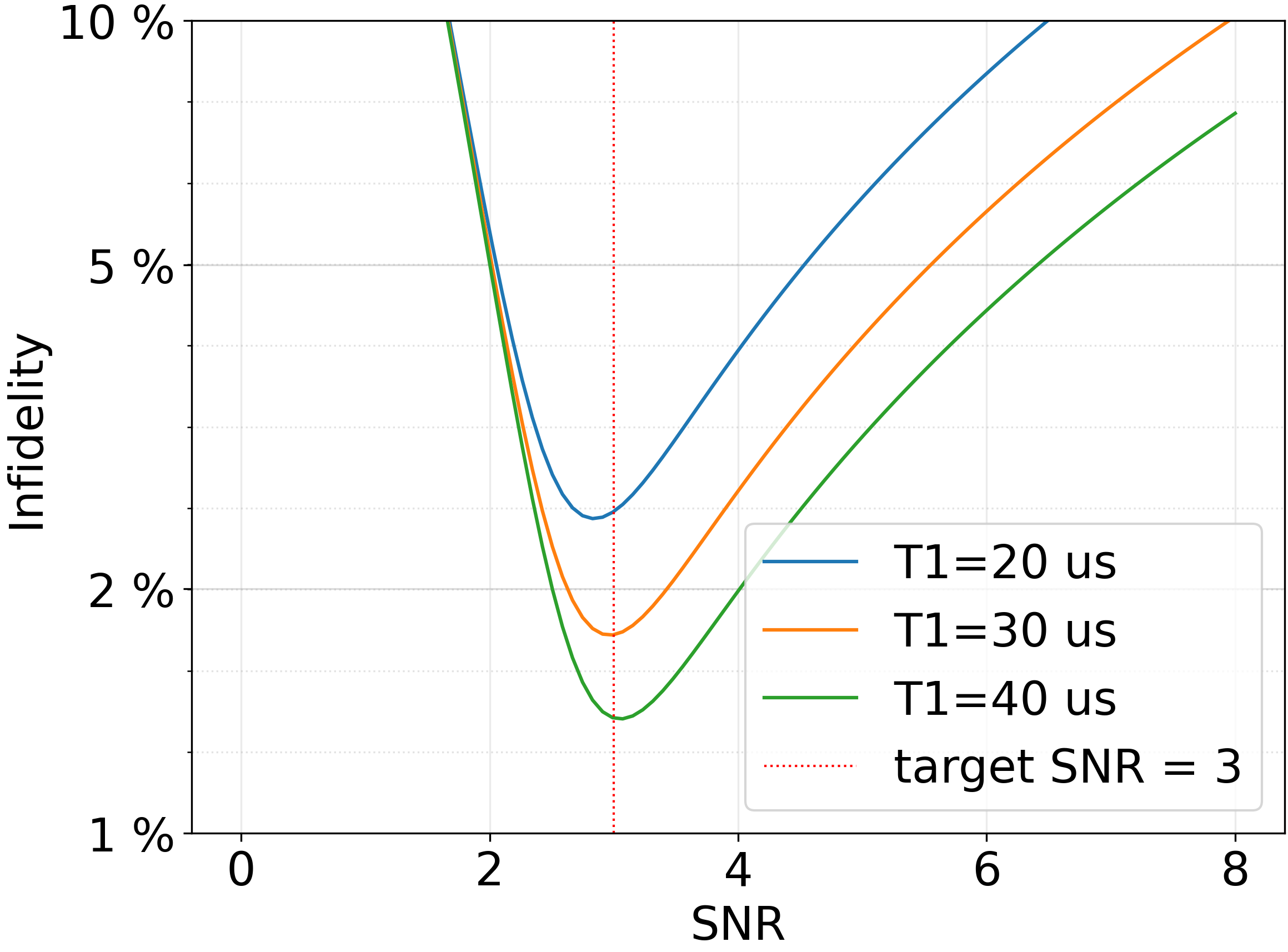}
  \caption{Estimated readout infidelity, $1-\mathrm{Fidelity}$, computed from Eq.~\eqref{eq:assign_fidelity_model} as a function of $\mathrm{SNR}$ for three representative relaxation times: $T_1=20\,\mu\mathrm{s}$ (blue), $30\,\mu\mathrm{s}$ (orange), and $40\,\mu\mathrm{s}$ (green). The vertical red dashed line marks the target $\mathrm{SNR}=3$. In our measurements, $T_1$ is typically $\sim 30\,\mu\mathrm{s}$ and can fluctuate to $\sim 40\,\mu\mathrm{s}$; we include $T_1=20\,\mu\mathrm{s}$ as a conservative lower bound. Across $T_1=20$--$40\,\mu\mathrm{s}$, operating at $\mathrm{SNR}=3$ yields an infidelity within $0.05\%$ of the minimum value, motivating our choice of target $\mathrm{SNR}$.}
  \label{fig:fidelity}
\end{figure}

Using the standard dispersive-readout scaling $\mathrm{SNR}\propto\sqrt{T_{\mathrm{RO}}}$, we eliminate $T_{\mathrm{RO}}$ in Eq.~\eqref{eq:assign_fidelity_model} to express the infidelity, $1-\mathrm{Fidelity}$, as a function of $\mathrm{SNR}$ and $T_1$ only. Figure~\ref{fig:fidelity} shows that increasing $\mathrm{SNR}$ reduces the infidelity up to $\mathrm{SNR}\approx 3$, while further increases yield diminishing returns because relaxation during the readout dominates. Based on this plateau behavior over the relevant range of $T_1$, we set the target to $\mathrm{SNR}=3$ and, throughout the main text, minimize $T_{\mathrm{RO}}$ subject to this fixed target.

\FloatBarrier
\FloatBarrier

\bibliographystyle{aipnum4-1}
\bibliography{aipsamp}

\end{document}